\documentclass[review,number,sort&compress]{elsarticle}
\usepackage{hyperref,amssymb}

\journal{Nucl. Instr. Meth. A}

\begin{document}

\begin{frontmatter}

\title{Operation of a double-sided CMOS pixelated detector at a high intensity $e^+e^-$ particle collider}

\author[iphc]{D.~Cuesta}
\author[iphc]{J.~Baudot}
\author[iphc]{G.~Claus}
\author[iphc]{M.~Goffe}
\author[iphc]{K.~Jaaskelainen}
\author[iphc]{L.~Santelj\fnref{myfootnote}}
\author[iphc]{M.~Specht}
\author[iphc]{M.~Szelezniak}
\author[iphc]{I.~Ripp-Baudot\corref{corr}}

\address[iphc]{Universit\'e de Strasbourg, CNRS, IPHC UMR 7178,  Strasbourg, France}

\cortext[corr]{Corresponding author}
\ead{isabelle.ripp@iphc.cnrs.fr} 
\fntext[myfootnote]{Present address: J. Stefan Institute, 1000 Ljubljana, Slovenia}

\begin{abstract}
This article reports the first operation of a double-sided CMOS pixelated ladder in a collider experiment, namely in the inner tracker volume of the Belle~II experiment during the Phase 2 run of the SuperKEKB collider. Design and integration of the detector system in the experiment interaction region is first described. The two modules operated almost continuously during slightly more than four months, recording data for the monitoring of the hit rate close to beams. Details of the off-line data analysis are provided and a method to estimate particle momentum from the 2 hits measured per crossing particle is proposed.
\end{abstract}


\end{frontmatter}

\section{Introduction}
\label{sec:introduction}
Tracking detectors in high energy particle physics experiments at colliders must sustain very high detector occupancy, e.g. produced by pile-up or beam-induced background particles. In the meantime, accurate particle tracking requires granular and light detection layers. In this respect, one of the most promising technologies for charged particle tracking in current experiments and their future upgrades is based on pixelated CMOS silicon sensors.

The PLUME detector (Pixelated Ladder with Ultra-low Material Embedding) \cite{bib:PLUMEref} is an ultra-light, double-sided ladder equipped with CMOS pixel sensors on both sides. This concept was developed originally in the perspective of a future ILD vertex detector within the collaboration of three institutes (DESY-Hamburg, IPHC-Strasbourg and the University of Bristol). It has already been characterized with various test beams, pions at CERN and electrons at DESY \cite{bib:boitrelle,bib:maria}. We report in this paper the first operation of a PLUME detector in the interaction region of a particle collider. 

Based on the PLUME concept, we developed and installed temporarily into the Belle~II detector \cite{bib:BelleII} a dedicated system, henceforth abbreviated as PLUME, with the ability to map hit rates in specific areas, while adding minimal material. As part of a wider ensemble of radiation monitors, that constituted the setup replacing the vertex detector during the Phase~2 run of the $e^+e^-$  SuperKEKB collider, PLUME contributed to monitoring the background induced by SuperKEKB in the inner tracker volume. 

The first run of SuperKEKB is called Phase 1 and took place in 2016. Each beam was circulated individually, no collisions were performed, and single beam backgrounds were studied by a dedicated setup called BEAST~II Phase 1 \cite{bib:beast2},  installed in the interaction region in place of the Belle~II detector. First collisions were delivered during the second run of SuperKEKB, a.k.a. Phase 2 run, that occurred from March to July 2018. For the Phase 2 run, the Belle~II detector was placed at the collision point and commissioned, and the final focus magnet was in place. However only one sector in the $\phi$ acceptance of the Belle~II vertex detector was installed, the rest of the inner tracker volume being equipped with  dedicated BEAST~II Phase 2 detectors \cite{bib:beast2VXD}. The main goal of the BEAST~II effort was to study and minimise the parasitic particle rate produced by the collider during the machine parameter tuning, to validate the simulation of this background and to check whether it was safe to install the full inner tracker in Belle~II.

This paper details the design, integration, operation experience and data analysis procedure of the PLUME system in BEAST~II during the Phase 2 run. Results on the background induced by SuperKEKB beams are not the goal of this report and will be presented in a dedicated paper in preparation in the Belle~II machine background group. 

Sections are organised as follows. The PLUME device and its acquisition system are described in section \ref{sec:system}. Data taking and detector stability during operations are discussed in section \ref{sec:data}. Section \ref{sec:doublesided} reviews the data analysis to extract hit rates and information related to double-sided measurement, before concluding in section \ref{sec:conclusion}.

\section{System description}
\label{sec:system}

\subsection{Detector design and geometry}
\begin{figure}[ht]
\centering
\includegraphics[width=12cm]{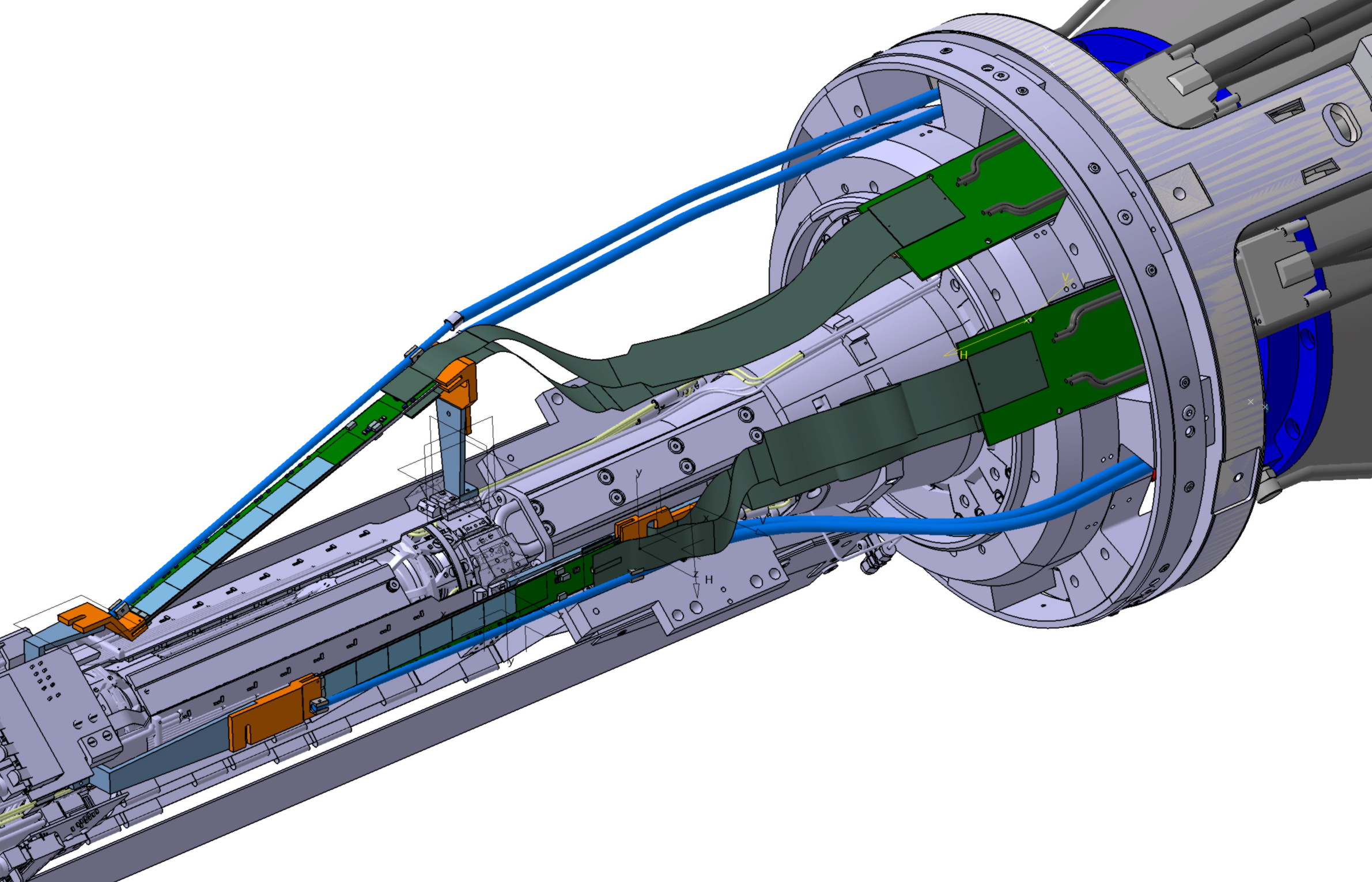}
\caption{CAD drawing of the two PLUME ladders around the SuperKEKB beam pipe in the interaction region.}
\label{fig:cad}
\end{figure}

\begin{figure}[ht]
\centering
\includegraphics[width=12cm]{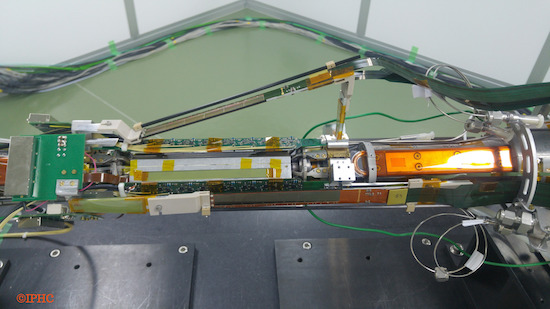}
\caption{Picture of the two PLUME ladders around the SuperKEKB beam pipe in the interaction region, before insertion in the Belle~II inner tracker volume.}
\label{fig:geometry}
\end{figure}

The PLUME detector in the Phase 2 setup of BEAST~II  consists of two 
double-sided modules shaped as ladders, each one equipped with 12 MIMOSA-26 sensors \cite{bib:mimosa26}. One ladder is installed parallel to the beam axis, at a radius $R$ = 5 cm and an azimuthal angle $\phi$ = 225$^{\circ}$ in the transverse plane around the beam pipe. The second ladder, located at $\phi$ = 135$^{\circ}$, is inclined with a polar angle $\theta$ = 18$^{\circ}$ with respect to the beam axis, and provides radial coverage from R = 5 cm to R = 9 cm, in order to scan the rate of background particles at roughly the same radii as the silicon strip vertex detector of Belle~II before its installation for the Phase 3 run in 2019. The PLUME layout in the Phase 2 setup of the BEAST~II inner tracker volume is shown in figures \ref{fig:cad} and \ref{fig:geometry}. In the following, the so-called outer or inner sides of a ladder means the ladder side facing respectively the Belle~II outer part of the detector or the beam.

\begin{figure}[ht]
\centering
\includegraphics[width=\linewidth]{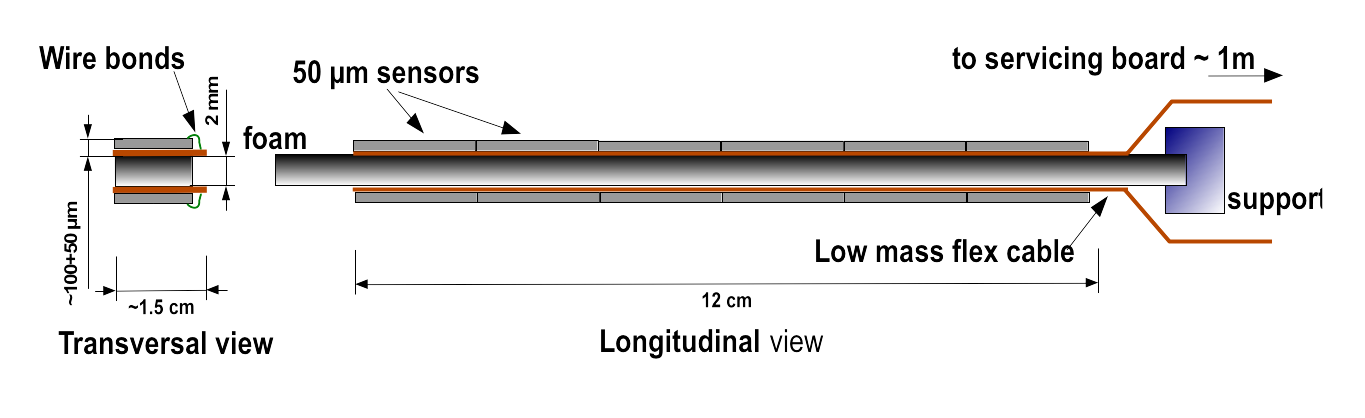}
\caption{Schematic concept of a PLUME ladder. Note that the vertical  and horizontal axes do not share the same scale.}
\label{fig:ladderlayout}
\end{figure}

\begin{figure}[ht]
\centering
\includegraphics[width=10cm]{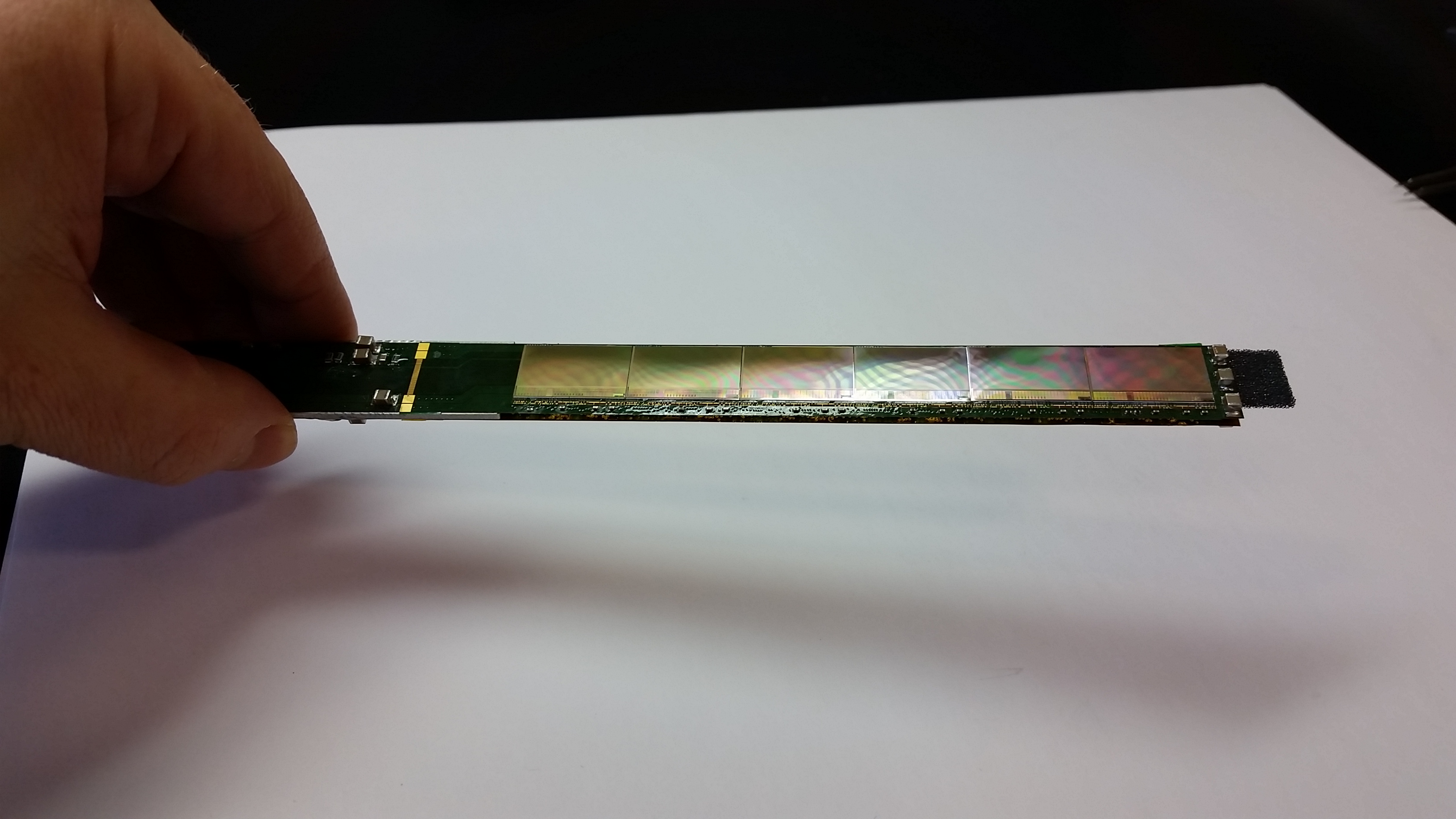}
\caption{Picture of a self-stiffened PLUME ladder (courtesy of J. Goldstein, university of Bristol).}
\label{fig:ladderphoto}
\end{figure}

Both ladders are self-stiffened thanks to the sandwich-effect of silicon sensors glued on a low-mass kapton-copper flexible PCB assembled on both sides of a low density SiC foam of thickness 2 mm (see figures \ref{fig:ladderlayout} and \ref{fig:ladderphoto}). Each ladder is very light with a total weight of about 7 g, resulting in a material budget of 0.42 \% of $X_0$. It is supported mechanically on one extremity in a cantilever way,  with halogen-free support structures made of polyetheretherketone (PEEK) and aluminum, screwed to the Belle~II inner tracker support structure. Passive radiochromic films \cite{bib:radiochromic} are glued on both sides of the ladder aluminium supports, to estimate the total radiation dose that impacted PLUME during the full Phase 2 run.
Both sides of the ladders are cooled down by blowing nitrogen gas through carbon tubes bored with 12 holes of diameter 800 $\mu$m facing the ladders. These carbon tubes are positioned alongside the edge of the ladder and the nitrogen flux of 4 l/min is shared between both sides of the ladder. 
With cooling, the sensor temperature varies from 40 $^{\circ}$C
 to 64  $^{\circ}$C.

\subsection{MIMOSA-26 CMOS sensors}

MIMOSA-26 sensors \cite{bib:mimosa26} have been designed in on the AMS 0.35 $\mu$m OPTO process. They are thinned down to a thickness of 50 $\mu$m. They feature a sensitive surface of 1$\times$2 cm$^2$, segmented in 576$\times$1,152 square monolithic active pixels of dimension 18.4$\times$18.4 $\mu$m$^2$. The pixel matrix is read out in a column-parallel rolling shutter mode. Analog signal of each pixel is converted to a binary digital output at the end of each of the 1152 columns of the pixel matrix, and passes through a 0-suppression stage with a discrimination threshold set as a multiple of the average measured pixel noise $\sigma$. This 0-suppression mechanism reduces by a factor 10 to 100 (depending on the occupancy level) the data flow produced by the 8$\times$10$^6$~pixels of each ladder, while maintaining a detection efficiency above 99.5~\% and a pixel fake rate of about or below $\times$10$^{-6}$ per read-out frame (for a discrimination threshold set to 8$\sigma$  and as long as the sensor temperature remains below 70$^{\circ}$C). Each pixel line is read out successively in about 200 ns, resulting in a total read-out time, also the pixel integration time, of 115.2~$\mu$s  for the full sensor matrix consisting of 576~pixel lines. This corresponds to a readout frequency of 8680~frames/s, and a capability to read out at least 10$^6$~hits/cm$^2$/s (depending on the hit spatial distribution). 

\subsection{Data acquisition system}
Beam background generated by SuperKEKB during Phase 2 was detected by BEAST~II detectors and registered by each detector's standalone DAQ (data acquisition) system, independently of the Belle~II global DAQ system including all Belle~II detectors. A synchronisation clock was provided by SuperKEKB and included in PLUME data. The PLUME detector during Phase 2 provided two kinds of outputs:\\
- EPICS output \cite{bib:EPICS} used to monitor on-line the beam-induced background and provide Hz-level feedback during the beam parameter fine tuning performed to increase the luminosity. \\
- ROOT output \cite{bib:Root} 
used for off-line studies of beam-induced backgrounds and their dependency on the beam conditions.

The PLUME DAQ system is based on two National Instrument Flex RIO PXIe 7962R acquisition boards equipped  with LVDS (Low-Voltage Differential Signaling)  NI-6585 front-end modules, hosted in a PXIe crate controlled by a PXIe 8135 CPU board. 
PLUME ladders generate a data flow that can reach at maximum 480 MB/s over 48 LVDS links at 80 Mb/s, acquired by the Flex RIO board firmware developed in VHDL and LabVIEW FPGA. 
The DAQ software written in LabVIEW and C languages, processes this data flow and routes it either to EPICS for on-line monitoring or to a RAID disk for off-line analysis.

\begin{figure}[ht]
\centering
\includegraphics[width=\linewidth]{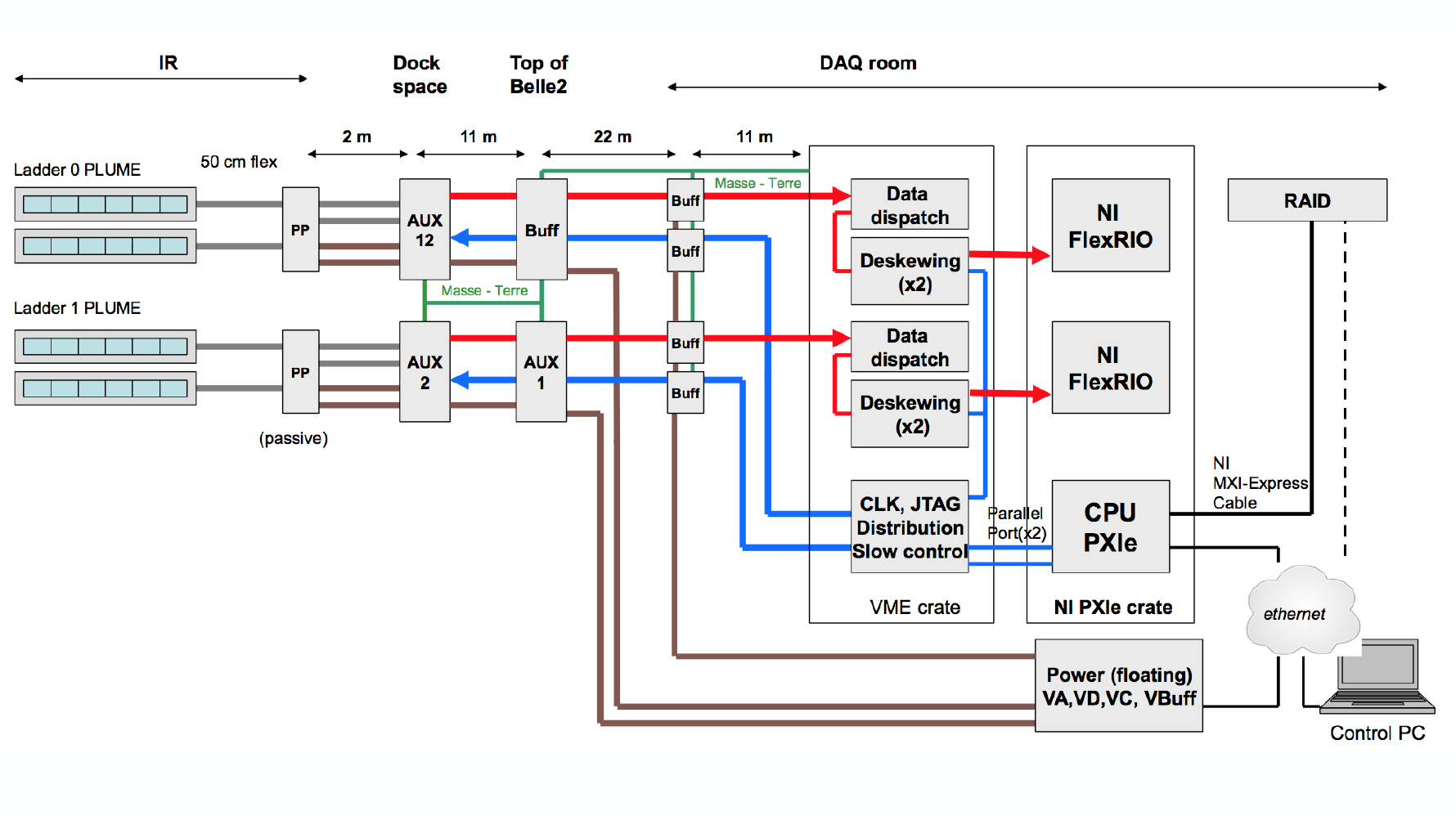}
\caption{Layout of the acquisition system of the 2 double-sided  PLUME ladders in the Phase 2 setup of BEAST~II vertex detectors, with indication of  the location of the different components. IR stands for Interaction Region.}
\label{fig:system}
\end{figure}

This standalone DAQ system is based on the one commonly used to read out the EUDET beam telescope \cite{bib:EUDET} equipped with MIMOSA-26 sensors, adapted here to particular integration constraints and also to read out 4 times more sensors.
In particular, two acquisition paths were designed, each one handling one ladder, see figure \ref{fig:system}. This design was developed to accommodate orthogonal constraints of electronic noise and radiation levels. The first DAQ path, referred to as AUX1-2 in figure \ref{fig:system}, has active components in a possible high radiation environment but allows to limit the electronic noise because the signal is buffered close to the detector, at 2.5 m from it. The second acquisition path is used for the second ladder and referred to as AUX1-AUX2 in figure \ref{fig:system}. In that case only radiation-tolerant power supply and slow control boards (AUX2) are placed 2.5 m away from the detector. The signal is buffered with AUX1 board, 13.5 m away from the PLUME detector on top of the Belle~II detector, in an environment less exposed to radiations. 
To reduce electronic noise and dispersion  of the propagation delay arising along the long transfer path (46 m between the detector and the electronic-hut), the 48  signal data are transmitted over differential LVDS pairs. Signal is buffered and re-amplified every 11 m (constrained by the cable length) and deskewed at the DAQ system level.
 In the end, the full system, including slow control of the detector, monitoring of the sensor temperature and power dissipation, fuse-box and protection against latch-up, consists of 13 custom-designed boards and is summarised in figure \ref{fig:system}.

\section{Data taking and detector stability}
\label{sec:data}

\subsection{Data analysed}
SuperKEKB started circulating the first beam during Phase 2 on 19 March 2018. Very first SuperKEKB collisions were delivered on 26 April and the Phase 2 run was completed on 17 July 2018. 
SuperKEKB being an asymmetric machine, there are two storage rings, called the High Energy Ring (HER, in which 7 GeV electrons are circulated) and the Low Energy Ring (LER, in which 4 GeV positrons are circulated). HER and LER beam parameters can be varied independently. During the Phase 2 run, the beam-induced background was carefully studied for the two beams in the following conditions: HER and LER beam current scans (bunch current and number of bunches), HER and LER beam size scans, luminosity scans, collimator settings, continuous injection,  beam steering scans, vacuum scrubbing. During these studies, the $\beta_y^*$ parameter was decreased from 8 to 3 mm (corresponding to a vertical beam size $\sigma_y$ of about 300 nm) and beam currents were increased up to nearly 800 mA in both LER and HER, to increase the instantaneous luminosity up to 0.54$\times$10$^{34}$ cm$^{-2}$ s$^{-1}$\cite{bib:Accel}. A number of runs were also operated under stable beam conditions to accumulate statistics for physics analysis.

PLUME was smoothly running and registering data during about 99 \% of the Phase~2 period. It was only seldom switched off manually, mainly during a few early vacuum scrubbing periods. Such conditions indeed generate possibly high radiation dose, which could have damaged PLUME sensors and compromised the quality of future data taking. Another reason to switch off PLUME was when the sensor temperature was increasing above 70 $^{\circ}$C because the cooling system of the BEAST~II vertex detector volume was switched off. Besides these very few intentional down-times, few spontaneous switch-offs were triggered by the security system of PLUME against single event latch-up. Such  over-consumption events occurred only a few times over 4 months of continuous running, at the level of the digital voltage, and were mainly generated during high beam injection noise as discussed in section \ref{sec:injection}.

\begin{figure}[ht]
\centering
\includegraphics[width=\linewidth]{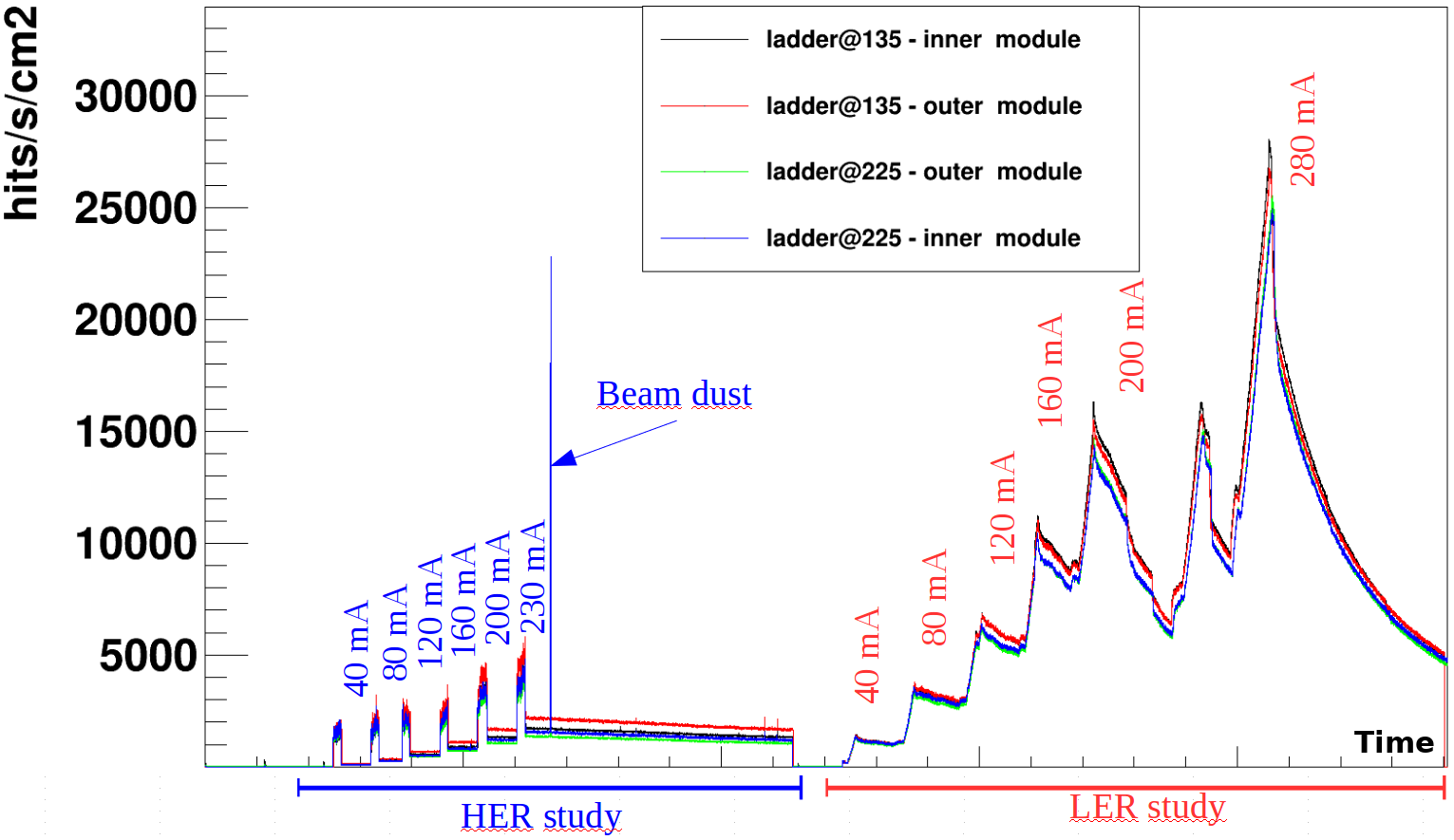}
\caption{Hit rate measured with the two PLUME ladders (inner and outer sides) as a function of time, for different LER and HER beam conditions.}
\label{fig:hitrate}
\end{figure}

As explained in section \ref{sec:system}, PLUME was operated permanently as a continuous hit rate monitor sending hit rates on-line without raw data saving, and sometimes also with full raw data storage for offline analysis allowing detailed studies of background rates as a function of beam parameters, as illustrated with figure \ref{fig:hitrate}. The analysis technique to extract hit rates is described in section \ref{sec:doublesided}.

\subsection{Radiation tolerance}
The SuperKEKB Phase~2 run was an important 
commissioning period that required tuning of multiple accelerator parameters and was exposing the  detector systems to possible direct beam losses.
A wide range of hit rates was expected in the detectors and no robust estimates of the expected total ionising dose and of the fluence of particles inducing non-ionising energy loss (NIEL) were available beforehand. 

The pixel fake rate of individual PLUME sensors was carefully monitored over the entire Phase 2 period in order to detect potential performance degradation induced by radiations.  
Radiations may indeed impact CMOS sensors in two ways. On the one hand, NIEL may damage the silicon crystalline structure, creating charge traps which  decrease the signal amplitude, and increasing the leakage current. On the other hand, cumulative ionising dose may lead to the accumulation of electric charges in oxide layers of CMOS circuits while they are powered up \cite{bib:RadHard_1,bib:RadHard_2}, generating additional temporal noise. Both effects may increase the pixel fake rate of  sensors like MIMOSA-26, that apply a signal threshold to discriminate between pixels registering an ionising particle crossing and those registering only electronic noise.

Radiation tolerance of MIMOSA-26 sensors equipping PLUME has been formerly studied with a 120 GeV $\pi$ beam at the CERN-SPS after exposing these circuits to 10 keV X-rays and/or to neutrons produced by a nuclear reactor \cite{bib:RadHard_1,bib:CmosTest}. Stable performances have been observed up to 1.5~kGy of total ionising dose combined with a fluence of $3\times10^{12}$~n$_{eq}$~cm$^{-2}$. However, after a total ionising dose of 3~kGy, the pixel fake rate significantly increased by a factor $\times3$ to $\times5$, depending on the discriminating threshold applied on sensors.

During Phase 2 operations, the non-ionising particle fluence was not expected to be an issue. Ionising radiation damage, on the other hand, could degrade the detector performance (depending on the level), but if not beyond the acceptable limits its effects can be mitigated either by increasing the discriminating threshold, or by masking the pixel using a programmable mask built into the sensor, or by masking the pixel off-line.

\begin{figure}[ht]
\centering
\includegraphics[width=\linewidth]{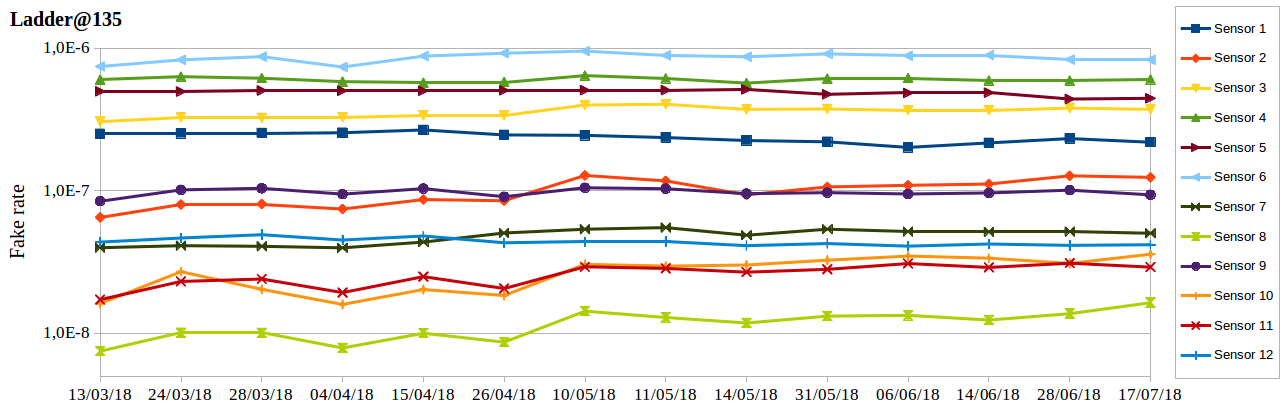}
\includegraphics[width=\linewidth]{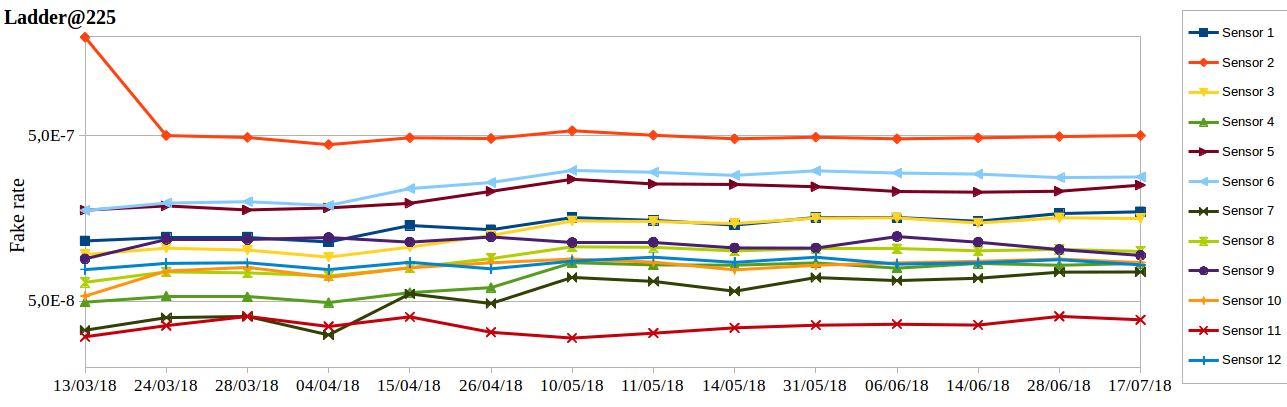}
\caption{Pixel fake rate measured per CMOS sensors as a function of time during the whole Phase 2 run.}
\label{fig:FakeRate}
\end{figure}

The average pixel fake rate over all PLUME sensors during the full Phase 2 run was measured to be 6$\times$10$^{-8}$ per pixel and per read-out frame with a discriminating threshold of 10$\sigma$.
It decreased to 5$\times$10$^{-10}$ by masking off-line 0.06 \% of the pixels, and to 2.5$\times$10$^{-10}$ by masking 0.1 \% of them.\\ 
PLUME data without beam were registered at regular intervals to  monitor the pixel fake  rate. No significant impact was observed, as shown in figure \ref{fig:FakeRate}, and the total integrated dose was estimated to be 200 Gy at maximum according to reference studies reported above in this section. This conclusion was cross-checked by radiochromic film measurements (see section \ref{sec:system}), which reported observed  radiation levels in the vicinity of PLUME ladders, decreasing with increasing radius from about less than 1 kGy to a few 100 Gy. 

\section{Extracting detector hit rates}
\label{sec:doublesided}

\subsection{Event selection}\label{sec:injection}
\begin{figure}[ht]
\centering
\includegraphics[width=\linewidth,width=10cm]{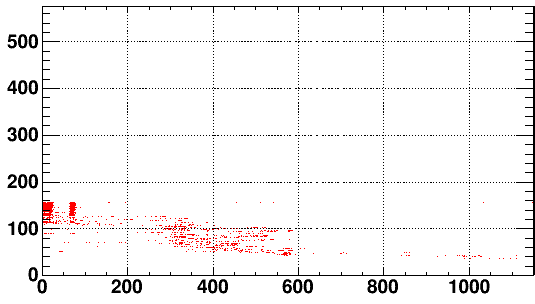}
\caption{A typical saturated MIMOSA-26 frame due to too high hit rate, showing a distinct pattern where only the first rows and columns can register fired pixels. The x and y axes correspond respectively to columns and lines of the sensor pixel matrix and each red point is a fired pixel.}
\label{fig:saturation}
\end{figure}

As already mentioned, SuperKEKB beam parameters were optimized during the Phase 2 run, resulting in a wide variation of the hit rate observed in PLUME.
Nevertheless, outside the beam injection periods, the detector occupancy remained below the MIMOSA-26 sensor limits. 
However when new bunches were injected in the beam, 
their initial non-optimal orbit led to a background rate increase by at least one order of magnitude. In such a situation, saturation was observed in the sensor due to the limited size of the embedded memory as explained in \cite{bib:mimosa26}. Figure \ref{fig:saturation} illustrates a real MIMOSA-26 frame with clearly a saturated event, while usual frames only feature a number of isolated impacts.

The hit rate cannot be estimated robustly in case of a saturated frame. Such frames can nevertheless  be easily removed from further analysis, based on a simple cut considering the number of fired pixels within the frame. This cut does not impact the measured hit rate since 
the injection frequency during Phase~2 was generally negligible with respect to the 1 Hz hit rate update, corresponding to about $10^4$ MIMOSA-26 frames. Additionally, this cut was not used for the specific study of the background conditions generated by newly injected bunches.

\subsection{Hit reconstruction}

Clusters (or hits) are reconstructed off-line from the list of fired pixels, i.e. showing a signal above the discrimination threshold, provided by sensors. The clustering algorithm gathers fired pixels with adjacent sides (not corners) and allows single-pixel hit.

\begin{figure}[ht]
\centering
\includegraphics[width=0.9\linewidth]{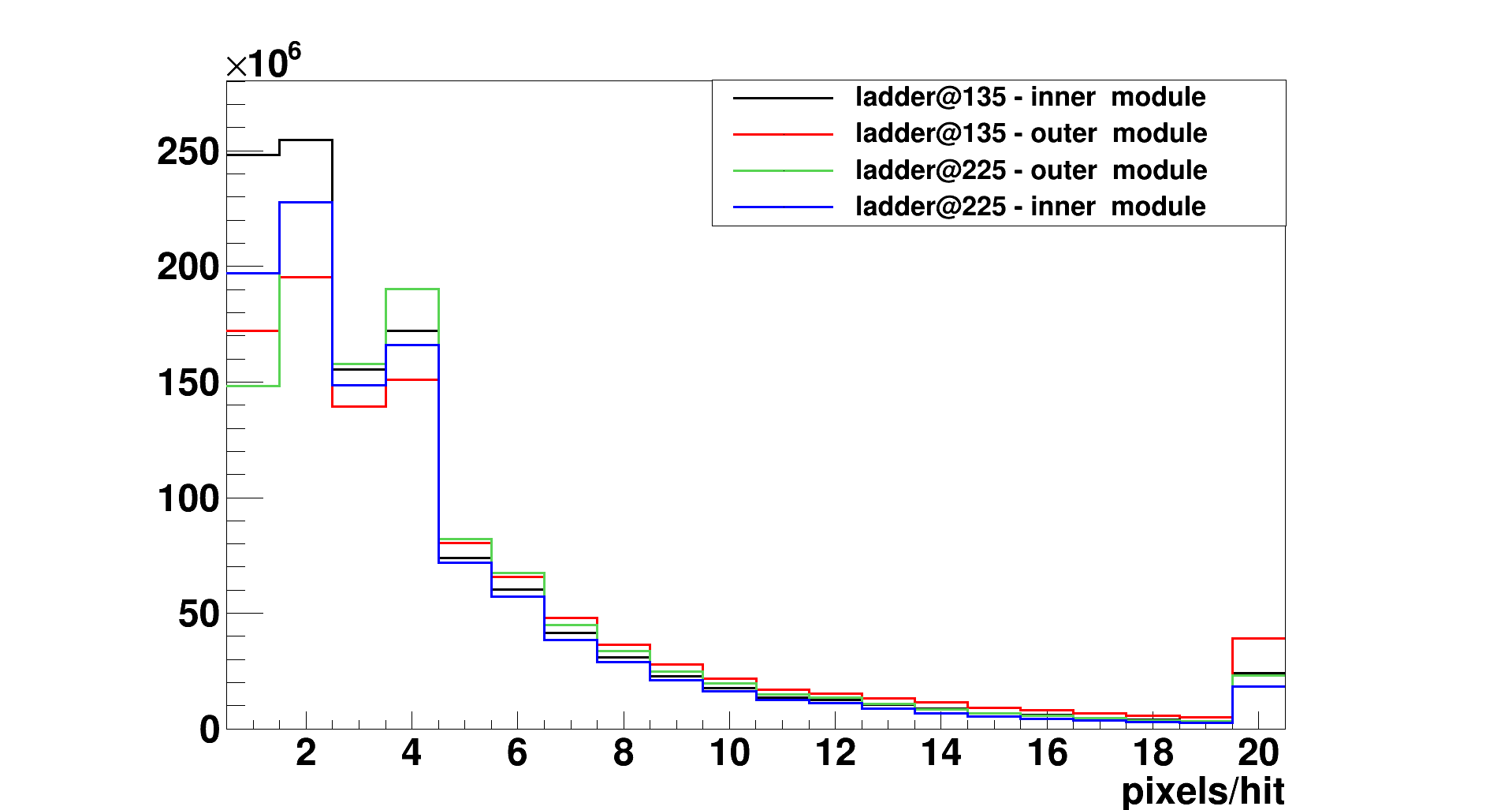}
\caption{Distributions of the number of pixels per cluster for the four PLUME modules, without relative normalization. The last bin contains all entries beyond 20 pixels.}
\label{fig:pixelMultiplicity}
\end{figure}

From previous beam tests with 120 GeV $\pi$ beam, the average cluster size for perpendicular crossing particles was observed to be 2 pixels with a standard deviation of 1 pixel, operating the sensor with a threshold of 10$\sigma$. The corresponding measured spatial resolution was measured to be $3.6 \pm 0.1 \> \mu$m \cite{bib:Resol}.
Of course, particles with much lower energies and a wide range of incoming angles were expected from beam induced background. The observed distribution of the number of pixels per cluster is displayed in figure \ref{fig:pixelMultiplicity}, where the last bin contains all entries larger than 20. Though 2-pixel clusters dominate the distribution, the long tail indicates the presence of large incidence angles. The non-negligible amount of single-pixel clusters reflects the presence of both particles depositing a small ionizing charge in the sensor or generating ionization close to the collecting diode, and fake hits.

\begin{figure}[ht]
\centering
\includegraphics[width=\linewidth]{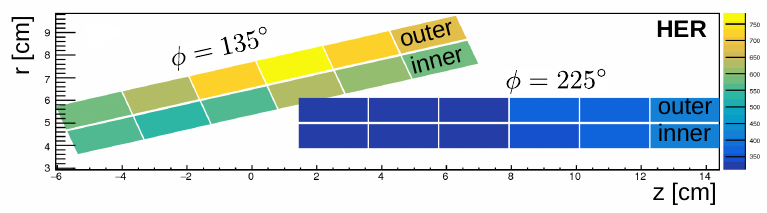}
\caption{Typical hit rate (arbitrary units) maps obtained with PLUME ladders as a function of radius r and z coordinate along the beam axis. Both ladders are represented, this means 6 sensors per side of each ladder.}
\label{fig:TandB}
\end{figure}

The position of each hit is computed from a simple average of the positions of all pixels associated to the hit. This position can potentially be used to detect background hot spots but usually a simple, sensor-based segmentation map was used, like depicted in figure \ref{fig:TandB}.

\subsection{Double-sided measurement}

The double-sided pixelated PLUME ladder provides two very close measured positions per crossing charged particle. As a first benefit, this feature brings redundancy in hit rate measurements between the two sides of the same ladder.
In addition, the two positions along each particle trajectory can be exploited for a helix fit to the track, assuming one additional constraint. This strategy can be applied for particles originating from the beam interaction point, which position is the needed third point. This is the case for background generated by beam-beam electromagnetic interactions, in contrast to background particles stemming from single beam effects that usually produce showers in surrounding material, which secondaries are detected by PLUME ladders. 

\begin{figure}[ht]
\centering
\includegraphics[width=0.7\linewidth]{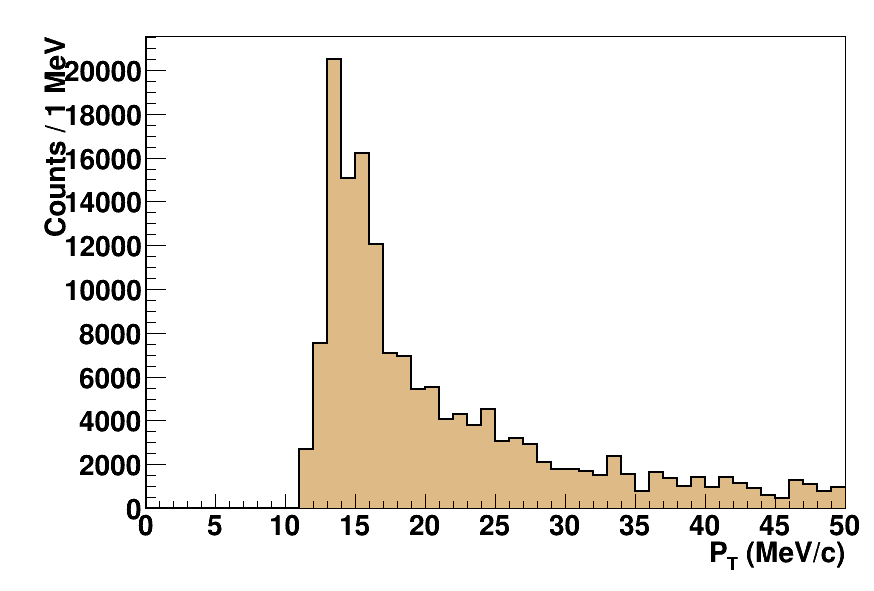}
\caption{Distribution of the reconstructed momentum with PLUME hits for particles emitted by simulated beam-beam interactions.}
\label{fig:LumiMomentum}
\end{figure}

\begin{figure}[ht]
\centering
\includegraphics[width=0.7\linewidth]{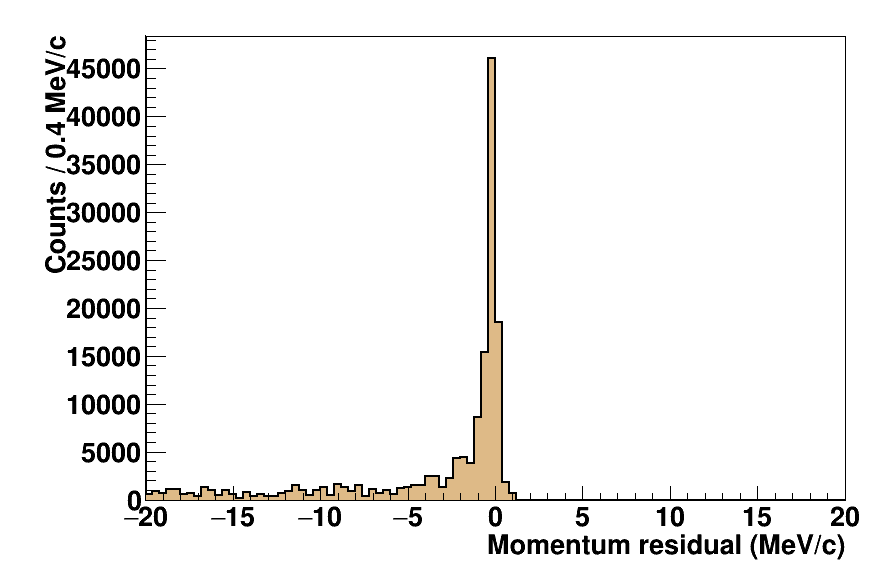}
\caption{Distribution of the momentum residual (true - reconstructed values) for particles emitted by simulated beam-beam interactions.}
\label{fig:LumiResidual}
\end{figure}

Figure~\ref{fig:LumiMomentum} displays the transverse momentum distribution obtained with such a constrained helix fit using two PLUME points and the interaction point, on simulated beam-beam interaction events. The distribution features a strong cut-off below 15~MeV/c, matching the minimal transverse momentum required to reach the PLUME ladder from the interaction point within the Belle~II 1.5 T magnetic field. The transverse momentum residual distribution depicted in figure~\ref{fig:LumiResidual} indicates that momenta are estimated within $\pm$1~MeV/c for 50~\% of the tracks. The long tail for the lower part of the residual corresponds to momentum over-estimation occurring when the origin of the particle is not the interaction point.

Such an analysis intended to discriminate background particles originating from beam-beam interactions, scaling with luminosity, against single-beam effects, using for instance the $\chi^2$ yielded by the track fit. This strategy however holds when hits from both PLUME sides can be associated with good efficiency and purity, which occurs for low hit rates resulting in low random hit probability.
However during the Phase 2 runs at the highest instantaneous luminosities, corresponding to the highest expected beam-beam background contributions, hit rates induced by single-beam background particles exceeded $50\times10^3$~hits/cm$^2$/s and were still 1,000 times higher than hit rates produced by beam-beam background particles.
Such rates led to an overwhelming combinatorial hit association and did not allow for proper use of the momentum fit ability offered by PLUME. 

\section{Conclusion}
\label{sec:conclusion}

We reported the first use of a detection system based on CMOS pixel sensors in the interaction region of an $e^+e^-$ collider, and also the first operation of double-sided pixelated detectors in a collider environment. Two double-sided PLUME ladders, previously developed for a future linear collider to feature the very low  material budget of 0.4~\%~$X_0$, were inserted temporarily in the inner volume of the Belle~II experiment. The read-out system was designed to match potential adverse radiation conditions and long data cable paths. Observed fake rates were equivalent to levels measured during beam tests under controlled conditions. No signs of radiation damaged were observed after the successful running period.
Both detecting ladders operated smoothly during the Phase~2 run of SuperKEKB and allowed monitoring online the rate of background particles generated by the beams.
Though simulation studies demonstrated the possibility to estimate the momentum of particles originating from the interaction point based on the double-sided feature of PLUME, the hit rate levels observed during the Phase 2 run was too high to allow a robust use of the method with these data.

\section*{Acknowledgement}
This work was supported by the Institut National de Physique Nucl\'eaire et de Physique des Particules of CNRS (CNRS/IN2P3) and by Investissements d'Avenir and Universit\'e de Strasbourg (IdEx grants W15RPE12 and W17RPD30). We thank Katsuro Nakamura (KEK laboratory, Tsukuba), for his help as our KEK liaison. We thank the SuperKEKB accelerator physicists. We are also grateful to Ingrid Gregor (DESY laboratory, Hamburg) and Joel Goldstein (university of Bristol) for helping manufactoring PLUME ladders for the Phase 2 setup of BEAST~II vertex detectors.

\section*{References}
\bibliography{mybibfile}

\begin{thebibliography}{10}
\expandafter\ifx\csname url\endcsname\relax
  \def\url#1{\texttt{#1}}\fi
\expandafter\ifx\csname urlprefix\endcsname\relax\def\urlprefix{URL }\fi
\expandafter\ifx\csname href\endcsname\relax
  \def\href#1#2{#2} \def\path#1{#1}\fi

\bibitem{bib:PLUMEref}
A.~Nomerotski, et~al., {PLUME collaboration: Ultra-light ladders for linear
  collider vertex detector}, Nucl. Instrum. Meth. A650 (2011) 208--212.
\newblock \href {http://dx.doi.org/10.1016/j.nima.2010.12.083}
  {\path{doi:10.1016/j.nima.2010.12.083}}.

\bibitem{bib:boitrelle}
B.~Boitrelle, {Development of a double-sided ladder for tracking in high-energy
  physics}, Ph.D. thesis, {University of Strasbourg} (February {2017}).

\bibitem{bib:maria}
R.~Maria, {Study of the time-dependent {\it CP} asymmetry in $D^0$ decays in
  the Belle~II experiment}, Ph.D. thesis, {University of Strasbourg} (October
  {2015}).

\bibitem{bib:BelleII}
T.~Abe, et~al., {Belle II Technical Design Report. }\href
  {http://arxiv.org/abs/1011.0352} {\path{arXiv:1011.0352}}.

\bibitem{bib:beast2}
P.~Lewis, et~al., {First Measurements of Beam Backgrounds at SuperKEKB}, Nucl.
  Instrum. Meth. A914 (2019) 69--144.
\newblock \href {http://dx.doi.org/10.1016/j.nima.2018.05.071}
  {\path{doi:10.1016/j.nima.2018.05.071}}.

\bibitem{bib:beast2VXD}
B.~Schwenker, et~al., {Operational experience and commissioning of the Belle II
  vertex detector}, PoS VERTEX2018 (2019) 006.
\newblock \href {http://dx.doi.org/10.22323/1.348.0006}
  {\path{doi:10.22323/1.348.0006}}.

\bibitem{bib:mimosa26}
C.~Hu-Guo, et~al., {First reticule size MAPS with digital output and integrated
  zero suppression for the EUDET-JRA1 beam telescope}, Nucl. Instrum. Meth.
  A623 (2010) 480--482.
\newblock \href {http://dx.doi.org/10.1016/j.nima.2010.03.043}
  {\path{doi:10.1016/j.nima.2010.03.043}}.

\bibitem{bib:radiochromic}
L.~Campajola, et~al., {Absolute dose calibration of {EBT}3 Gafchromic films},
  Journal of Instrumentation 12~(08) (2017) P08015--P08015.
\newblock \href {http://dx.doi.org/10.1088/1748-0221/12/08/p08015}
  {\path{doi:10.1088/1748-0221/12/08/p08015}}.

\bibitem{bib:EPICS}
\href{https://epics-controls.org}{{Web page of EPICS, The Experimental Physics
  and Industrial Control System, }}(accessed in December 2019).
\newline\urlprefix\url{https://epics-controls.org}

\bibitem{bib:Root}
\href{https://root.cern.ch}{{Web page of the ROOT Data Analysis Framework,
  }}(accessed in December 2019).
\newline\urlprefix\url{https://root.cern.ch}

\bibitem{bib:EUDET}
\href{https://telescopes.desy.de}{{Web page of EUDET-type beam telescopes,
  }}(accessed in December 2019).
\newline\urlprefix\url{https://telescopes.desy.de}

\bibitem{bib:Accel}
Y.~Funakoshi, et~al., {Operation of SuperKEKB in Phase 2}, in: {Proceedings,
  62nd ICFA Advanced Beam Dynamics Workshop on High Luminosity Circular $e^+
  e^-$ Colliders (eeFACT2018), Hong Kong, China, September 24-27, 2018}.
\newblock \href {http://dx.doi.org/10.18429/JACoW-eeFACT2018-WEPBB01}
  {\path{doi:10.18429/JACoW-eeFACT2018-WEPBB01}}.

\bibitem{bib:RadHard_1}
M.~Deveaux, et~al., {Radiation Tolerance of CMOS Monolithic Active Pixel
  Sensors with Self-Biased Pixels}, Nucl. Instrum. Meth. A624 (2010) 428--431.
\newblock \href {http://arxiv.org/abs/0908.4202} {\path{arXiv:0908.4202}},
  \href {http://dx.doi.org/10.1016/j.nima.2010.04.045}
  {\path{doi:10.1016/j.nima.2010.04.045}}.

\bibitem{bib:RadHard_2}
M.~Garcia-Sciveres, N.~Wermes, {A review of advances in pixel detectors for
  experiments with high rate and radiation}, Rept. Prog. Phys. 81~(6) (2018)
  066101.
\newblock \href {http://arxiv.org/abs/1705.10150} {\path{arXiv:1705.10150}},
  \href {http://dx.doi.org/10.1088/1361-6633/aab064}
  {\path{doi:10.1088/1361-6633/aab064}}.

\bibitem{bib:CmosTest}
G.~Voutsinas, et~al., {Studies for a 10 $\mu$s, thin, high resolution CMOS
  pixel sensor for future vertex detectors}, Nucl. Phys. Proc. Suppl. 215
  (2011) 48--50.
\newblock \href {http://dx.doi.org/10.1016/j.nuclphysbps.2011.03.131}
  {\path{doi:10.1016/j.nuclphysbps.2011.03.131}}.

\bibitem{bib:Resol}
M.~Winter, et~al., {Development of CMOS Pixel Sensors fully adapted to the ILD
  Vertex Detector Requirements}, in: {International Workshop on Future Linear
  Colliders (LCWS12) Arlington, Texas, USA, October 22-26, 2012}, 2012.
\newblock \href {http://arxiv.org/abs/1203.3750} {\path{arXiv:1203.3750}}.

\end{thebibliography}

\end{document}